%% file: main.tex
\documentclass[sigconf]{acmart}

\usepackage{color}
\usepackage{amsmath,amsfonts}
\usepackage{graphicx}
\usepackage{xcolor}

\usepackage{url}            
\usepackage{nicefrac}       

\usepackage{diagbox}
\usepackage{multirow}

\newenvironment{myitemize}{\begin{list}{$\bullet$}
		{\setlength{\topsep}{1mm}
			\setlength{\itemsep}{0.25mm}
			\setlength{\parsep}{0.25mm}
			\setlength{\itemindent}{0mm}
			\setlength{\partopsep}{0mm}
			\setlength{\labelwidth}{15mm}
			\setlength{\leftmargin}{4mm}}}{\end{list}}

\AtBeginDocument{%
  \providecommand\BibTeX{{%
    \normalfont B\kern-0.5em{\scshape i\kern-0.25em b}\kern-0.8em\TeX}}}

\copyrightyear{2021}
\acmYear{2021}
\setcopyright{acmcopyright}\acmConference[Arxiv '21]{Arxiv preprint version}{November 17--18, 2021}{Arxiv}

\begin{document}

\title{Learning-based Framework for Sensor Fault-Tolerant Building HVAC Control with Model-assisted Learning}

\author{Shichao Xu}
\affiliation{%
  \institution{Northwestern University}
  \city{Evanston}
  \country{USA}}
\email{shichaoxu2023@u.northwestern.edu}

\author{Yangyang Fu}
\affiliation{%
  \institution{Texas A\&M University}
  \city{College Station}
  \country{USA}}
\email{yangyang.fu@tamu.edu}

\author{Yixuan Wang}
\affiliation{%
  \institution{Northwestern University}
  \city{Evanston}
  \country{USA}}
\email{yixuanwang2024@u.northwestern.edu}

\author{Zheng O'Neill}
\affiliation{%
  \institution{Texas A\&M University}
  \city{College Station}
  \country{USA}}
\email{zoneill@tamu.edu}

\author{Qi Zhu}
\affiliation{%
  \institution{Northwestern University}
  \city{Evanston}
  \country{USA}}
\email{qzhu@northwestern.edu}


\begin{abstract}
As people spend up to $87\%$ of their time indoors, intelligent Heating, Ventilation, and Air Conditioning (HVAC) systems in buildings are essential for maintaining occupant comfort and reducing energy consumption. These HVAC systems in smart buildings rely `on real-time sensor readings, which in practice often suffer from various faults and could also be vulnerable to malicious attacks. Such faulty sensor inputs may lead to the violation of indoor environment requirements (e.g., temperature, humidity, etc.) and the increase of energy consumption.
While many model-based approaches have been proposed in the literature for building HVAC control, it is costly to develop accurate physical models for ensuring their performance and even more challenging to address the impact of sensor faults. 
In this work, we present a novel learning-based framework for sensor fault-tolerant HVAC control, which includes three deep learning based components for 1) generating temperature proposals with the consideration of possible sensor faults, 2) selecting one of the proposals based on the assessment of their accuracy, and 3) applying reinforcement learning with the selected temperature proposal. Moreover, to address the challenge of training data insufficiency in building-related tasks, we propose a model-assisted learning method leveraging an abstract model of building physical dynamics. 
Through extensive experiments, we demonstrate that the proposed fault-tolerant HVAC control framework can significantly reduce building temperature violations under a variety of sensor fault patterns while maintaining energy efficiency.
\end{abstract}






\maketitle

\input{section/introduction}
\input{section/related_work}

\input{section/method}

\input{section/experiment}
\input{section/conclusion}

\bibliographystyle{ACM-Reference-Format}
\bibliography{reference}

\end{document}

%% file: section/introduction.tex
\section{Introduction}
\label{sec:intro}

People spend up to $87\%$ of their time in enclosed buildings nowadays~\cite{klepeis2001national}. 
As Heating, Ventilation, and Air-Conditioning (HVAC) systems control the indoor environment of buildings and have a significant impact on occupant comfort, productivity, and physical/mental health, it is important to ensure their performance and reliability. 
In these systems, sensors, in particular temperature sensors, play a vital role in collecting real-time environment condition and facilitating HVAC applications.
However, temperature sensors are not always in normal working condition, due to passive faults and active cyber-attacks.
Passive sensor faults such as sensor bias and sensor drifting over a long time contribute more than 25\% to the variable air volume (VAV) terminal unit faults~\cite{qin2005fault}.
Cyber-attacks on HVAC control systems (i.e., corruption of temperature sensor readings to affect critical control programs) are becoming possible due to increasing connectivity of buildings to external networks for supporting remote management and cloud-based analytics. For example, Building Automation and Control Networks (BACnet)~\cite{newman2013bacnet}, the most popular communication protocol for buildings, has been reported to have multiple vulnerabilities that can be used to launch cyber-attacks on building control systems~\cite{holmberg2003bacnet}.
Moreover, HVAC systems still need to provide services when under faults or attacks, as diagnosing the problems and fixing the sensors often takes a significant amount of time.
This highlights the increasing need for developing HVAC controls that can tolerate sensor faults and cyber-attacks and increase system resilience.

There are a number of works in the literature related to sensor fault-tolerant control for building energy systems. 
Ma and Wang~\cite{ma2012fault} proposed a fault-tolerant model predictive control strategy to provide resilient operation of a building chiller plant system under typical faults such as condenser water supply temperature sensor bias. 
Yang et. al.~\cite{yang2014optimum} presented an online fault-tolerant control strategy for fixed bias faults in the supply air temperature sensor. The sensor faults are detected by using a pre-trained support vector regression (SVR) model.
\cite{gunes2015improving} employed a rule-based method (e.g., using sensor reading from the nearest zone) to mitigate the zone air temperature sensor reading spikes.
The work in~\cite{papadopoulos2018distributed} built a physical model for a multi-zone building and with zone air temperature sensor faults, and assumed that only one thermal zone would be affected by the sensor fault at a time.
Faults in sensors other than temperature sensors are also studied for tolerant control design. 
Wang et. al.~\cite{wang2002fault} applied a neural network model to detect and compensate outdoor air flow rate sensor faults, and provided a fault-tolerant control strategy to regain the control of outdoor air flow rate. 
However, the above literature has the following limitations: 1) simple assumptions in terms of fault occurrences are used: for instance, \cite{papadopoulos2018distributed} assumed that only one thermal zone would be affected by the zone air temperature sensor fault at a time, which is often not the case in practice; 2) studies were mostly designed for passive faults such as fixed sensor bias~\cite{ma2012fault,jin2006fault,yang2014optimum}, and might not be applied to active attacks that only last for a short duration but with high intensity; 3) significant efforts are required to obtain an accurate online state predictor, such as detailed physics-based models or SVR model, for fault detection in the fault-tolerant control. Therefore, how to provide resilient control for HVAC systems under abnormal sensor readings still remains an open challenge.

In this work, we develop a \emph{learning-based sensor fault-tolerant control framework} for building HVAC systems with novel deep neural network-based learning techniques.
Specifically, our framework includes three major components. 
First, as the raw sensor readings may be faulty, a  neural network-based temperature predictor is designed based on historical sensor data to provide an alternative estimation of the true temperature. 
Then, both temperature proposals (raw sensor reading and the temperature predictor output) are sent to a neural network-based selector, which assesses the two temperature proposals with consideration of the historical trend and selects one deemed more trustworthy. 
Finally, a deep reinforcement learning (DRL) based HVAC controller takes the chosen temperature as the current system state and applies control actuation. 
These learning-based techniques together provide a robust HVAC control framework that can maintain desired temperature and reduce energy consumption under sensor faults.




While our machine learning based techniques can remove the need for developing detailed and costly building physical models, they face their own challenges in training data availability. In particular, for a new building, we may have to wait for months to collect enough data for training the learning-based components. To address this challenge, we propose a \emph{model-assisted learning} approach that helps the learning components extract knowledge from an \emph{abstract physical model} and only requires a limited amount of additional labeled data collected from real buildings for training. There are a number of abstract physical models available in the literature~\cite{maasoumy2011model, toub2019model}. They require much less effort to develop than the accurate physical models (e.g., those used in EnergyPlus~\cite{crawley2000energy}). While they alone are often not accurate enough for building HVAC control, their capturing of the underlying physical laws can guide the learning process for the neural network-based components and significantly improve the learning effectiveness. 

To summarize, our work makes the following contributions:
\begin{itemize}
    \item We present a novel sensor fault-tolerant learning-based framework to achieve sensor fault resilience on building HVAC control. The framework includes three neural network-based components: a temperature predictor that estimates the true temperature, a selector that assesses the predictor output and the raw sensor reading and selects one, and a DRL-based controller that generates the control signal.
    \item We develop a novel learning method called model-assisted learning, which leverages the knowledge from an abstract physical model to enable learning with a small amount of labeled data. 
    \item We conduct a number of experiments on buildings with a single thermal zone and multiple zones, and demonstrate the effectiveness of our fault-tolerant framework under various types of sensor anomalies. We also highlight how model-assisted learning can improve the learning process and reduce the need for training data.
    
\end{itemize}

The rest of the paper is as follows. Section~\ref{related-work} discusses further about the related literature. Section~\ref{method} introduces
our approach, including the design of the sensor fault-tolerant framework and model-assisted learning. Section~\ref{experiment} shows the experiments and related ablation studies. Section~\ref{conclusion} concludes the paper.

%% file: section/related_work.tex
\section{Related Work}
\label{related-work}
\subsection{Building HVAC control}
Building HVAC supervisory controllers can be categorized into two groups, \emph{model-based controllers} and \emph{model-free controllers}. 
Classic model-based HVAC controllers are often developed based on fundamental physics laws (e.g., considering heat transfer and airflow balance). For example, \cite{maasoumy2011model} designed a hierarchical control algorithm based on modeling building thermal dynamics as an RC network, which uses resistance and capacitance elements to model the building envelope heat transfer. 
\cite{toub2019model} also used an RC network model and designed a model predictive control algorithm for minimizing the building energy consumption. There are other works~\cite{salakij2016model, xu2017model} that use similar abstract physical models.  
However, While being easy to develop and fast to run, these abstract physical models often suffer from inaccuracy. In contrast, detailed physical models such as EnergyPlus consider a variety of complex factors, including building layout, wall materials, light, shading, occupant behaviors, etc. They are much more accurate, but are typically hard to build and slow to run.  

Model-free HVAC controllers usually learn control strategies from historical data. In recent years, DRL-based methods have been explored in works such as~\cite{wei2017deep, zhang2018deep}, where techniques such as deep Q-learning (DQN) and asynchronous advantage actor-critic algorithms (A3C) are applied. Methods have also been proposed to learn DRL parameters by leveraging building simulation tools~\cite{gao2020deepcomfort, naug2019online, li2021towards, xu2020one}.
In this paper, we combine the strength of both model-free and model-based methods, by developing a learning-based framework with neural network-based components and leveraging abstract physical models to improve the learning process.
\subsection{Addressing sensor faults in buildings}
There has been a number of works in the literature addressing sensor faults in buildings. In~\cite{du2014sensor}, a fault detection method based on correlation analysis was proposed for detecting sensor bias or complete failure. \cite{du2014fault} proposed a neural network-based strategy with clustering analysis to detect sensor faults in the HVAC system and diagnose the sources. \cite{wang2005sensor} presented an online strategy based on the principal component analysis (PCA) to detect, diagnose and validate sensor faults in centrifugal chillers. More investigations can be found in \cite{liu2019sensor, mirnaghi2020fault, reppa2014distributed, da2012knowledge, fonollosa2013algorithmic, kim2018review}. However, these works focus on fault detection and diagnosis, not fault-tolerant control. 

There are some existing works for sensor fault-tolerant control in building energy systems, such as~\cite{ma2012fault, jin2006fault, yang2014optimum, gunes2015improving, papadopoulos2018distributed, wang2002fault}. For instance, Gunes et. al.~\cite{gunes2015improving} followed the model-based design paradigm and used rule-based methods to mitigate the negative effect of specific sensor faults. Papadopoulos et. al.~\cite{papadopoulos2018distributed} built a complex physical model for building, and designed a fault model based on the assumption that sensor faults occur in a single zone at each time. Jin and Du~\cite{jin2006fault} used principal component analysis, joint angle method and compensatory reconstruction to detect, isolate and reconstruct the fixed bias fault in supply air temperature sensors. However, as we outlined in the introduction, the above studies have significant limitations in the usage of simple or restricted assumptions, the focus on only passive faults with fixed sensor bias, and the need of significant efforts for obtaining an accurate online state predictor (e.g., with detailed physics-based models or SVR model). In contrast, our learning-based approach provides resilient control in broader and more practical cases.

\subsection{Learning with limited data and abstract physical model}
When dealing with a limited amount of labeled data in training, techniques such as weakly supervised learning~\cite{zhou2018brief, sun2010multi}
and semi-supervised learning~\cite{papandreou2015weakly, zhu2009introduction, chen2020simple} are often considered. However, in our case, even obtaining unlabeled data from real building operations could be a long process. Thus, we leverage the information from abstract physical models such as those in~\cite{maasoumy2011model, toub2019model}
to reduce the data needed for training. This approach is in principle related to model distillation techniques~\cite{hinton2015distilling, kim2016sequence} that distill the physical model into a neural network and then fine-tune the network with available labeled data. However, unlike in the case for those approaches (which focus on domains such as computer vision), there is not enough unlabeled data in the realistic data distribution that can be fed into the model for distillation in our problem. Thus, we propose model-assisted learning to overcome this difficulty, by leveraging abstract physical models to generate better initial points for model find-tuning.

%% file: section/method.tex
\section{Methodology}
\label{method}
\begin{figure*}[htb]
\centering
\includegraphics[width=17.6cm]{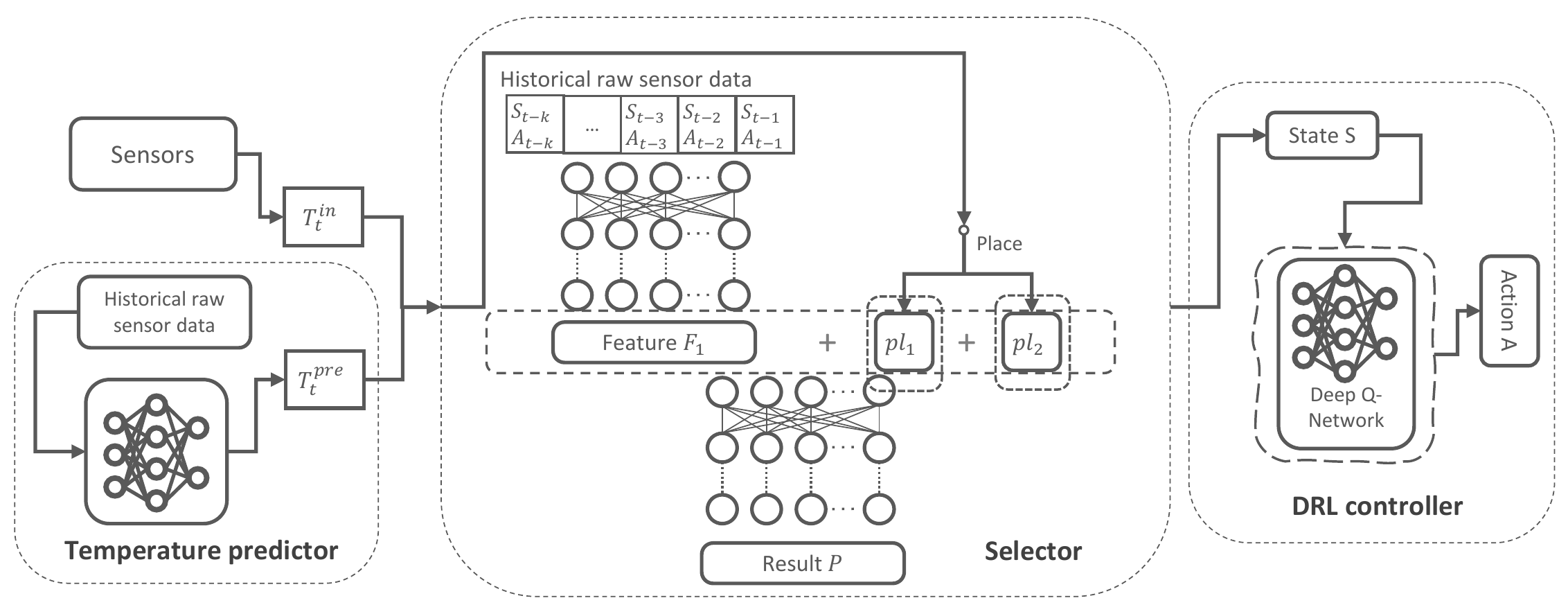}
\caption{Overview of our sensor fault-tolerant framework for building HAVC system. There are three main components: two modules providing temperature proposals on the left, a selector in the middle, a DQN-based HVAC controller on the right. The temperature proposals consist of the raw sensor reading $T^{in}_t$ and the current temperature prediction $T^{pre}_t$ that comes from the learned temperature predictor, which leverages the historical sensor data. The proposal selector provides a classification result to choose between the predictor output and the raw sensor value. Then, the DRL controller takes the selected temperature proposal and calculates the corresponding control action.}
\label{fig:overview}
\end{figure*}

\subsection{System model}
\label{system-model}
We adopted a multi-zone building model with the fan-coil system from~\cite{wei2017deep, xu2020one}, where there is a building with $n$ thermal zones, and a fan-coil system is equipped to provide the conditioned air at a given supply air temperature $T^{air}$ for each thermal zone. The airflow rate in each zone is chosen from multiple discrete levels $\{f_1, f_2, \cdots, f_m\}$, and corresponding to $m$ control actions $a_i$ for each zone $i$. With all $n$ thermal zones, the control action set is denoted as $A = \{a_1, a_2, \cdots, a_n\}$. In this paper, we denote the current physical time as $t$, the ambient temperature, indoor temperature for zone $i$, and the control action at time $t$ as $T^{out}_t$, $T^{in(i)}_t$, $A_t$, respectively, and we set $T^{in}_t = \{T^{in(i)}_t | i \in 1 \cdots n\}$. The system sends current states (indoor and ambient temperatures) to the HVAC system with a period of $\Delta t_s$ (which is the simulation period on building simulation platform), and the building HVAC controller provides the control signal (supply airflow rates) with a period of $\Delta t_c$ (i.e., the control period).

\subsection{Sensor fault-tolerant DRL framework}
\label{Sensor fault-tolerant DRL framework}
Fig.~\ref{fig:overview} depicts the overview of our sensor fault-tolerant DRL framework. It includes three parts:
the first part on the left is a neural network-based temperature predictor for providing an alternative estimation (rather than the raw sensor reading) of the indoor temperature, 
the second part in the middle is a proposal selector that assesses the temperature proposals from the raw sensor reading $T^{in}_t$ and the temperature prediction $T^{pre}_t$ and selects one, 
and the third part on the right is a DRL-based HVAC controller. 
With the design of the predictor and the selector, the DRL controller receives a refined temperature reading as part of its inputs and can maintain a stable performance against sensor faults or attacks. The details of each module are introduced in the following sections. Note that all the modules are trained individually and assembled into the framework after training. 

\subsubsection{Temperature predictor}
\label{temperature predictor}
The temperature predictor aims to provide a temperature prediction for the current temperature based on the historical sensor readings with possible faults and other system states. Note that we mark the current system state as $S_t$, where $S_t = (t, T^{in}_t, T^{out}_t)$.

Firstly, the temperature predictor is a neural network that consists of five fully-connected layers. Except for the last layer, all layers are filtered by a ReLU activation function, and all fully-connected layers are sequentially connected (detailed neuron number settings can be found later in Table~\ref{hyperparam} of Section~\ref{experiment}). In the test stage of the temperature predictor, 
the network takes the historical states aligned with the historical control actions (airflow rate) as the data inputs at time $t$
, and then outputs a current temperature prediction value $T^{pre}_t$. 

The training data for the predictor network is collected by running a straightforward ON-OFF controller on the building HVAC system for several days (in experiments we use 8 days). For new buildings, this could be done during the first several days of their operation, in which case we may assume that the data collected over this short period of time has not been polluted by sensors faults or attacks. 
And we get a (state, action) sequence from $(S_{1}, A_{1})$ to $(S_{L}, A_{L})$. For the convenience of supervised training, we select data sequences 
\begin{equation*}
    \{ \langle ( S_{t-k}, A_{t-k} ), ( S_{t-k+1}, A_{t-k+1} ), \cdots, ( S_{t-1}, A_{t-1} )\\  \rangle\}
\end{equation*}
with length $k$ and $t \in [k+1, L]$ from the historical data. These sequences are chosen with an interval $v$, which means that $t \in [k+1, L]$ is selected in the format $k+1, k+v+1, k+2v+1, \cdots$. The collected data set is used as the training data inputs of the neural network, with the corresponding label $S_t$ for each data sequence. Then, we train the neural network based on the loss function $\mathcal{L}_{pre}$ as
\begin{equation}
    \mathcal{L}_{pre} = \parallel (T^{pre}_t + T^{pre}_{ofs}) - T_t \parallel ^2,
\end{equation}
where $T^{pre}_t$ is the temperature prediction at time $t$ from the network's output, $T^{pre}_{ofs}$ is an estimated offset for bringing the absolute mean value of the neural network's output close to zero, which lowers the difficulty for the neural network learning through the given data sequences (it is a fixed hyper-parameter; setting can be found in Table~\ref{hyperparam} later).
$T_t$ is the actual indoor temperature, which is the ground truth label. After finishing training, the predictor can take the historical system states containing the raw sensor reading to generate the temperature prediction. We should mention that these historical system states in the test stage may contain faulty sensor readings, so we also include some faulty sensor reading in the training data for temperature prediction.
The designing of this training strategy using historical data with slightly faults is inspired by our preliminary experiments, which indicated that adding slightly faulty sensor reading to the training data could increase the performance on temperature predictions, compared to training with non-faulty data or data with high frequency faulty data. In other words, for enhancing the robustness of the temperature prediction, we use the historical system states under the independent and identically distributed (IID) faults with occurring probability $P_{pre}$. IID faults here mean that the fault can happen at each individual simulation step with probability $P_{pre}$. If the fault occurs, it uniformly selects a random number from $[T^{out}_l, T^{out}_u]$, which is the upper and lower boundary of the ambient temperature, to replace the original sensor temperature reading. And the temperature predictor takes benefit from randomized faults in the reading, which leads to a more robust output.

\subsubsection{Temperature proposal selector}
\label{temperature proposal selector}
The temperature proposal selector aims to choose the best candidate from the temperature proposals and send it to the DRL controller for further control steps. We train this module in a self-supervised way, where all the training labels are generated automatically and the objective is to distinguish between the normal data and the faulty data. Apart from the comparison between the normal and faulty, we also make the comparison among the faulty data and indicate which one is closer to the actual temperature value. This extra comparison further boosts the proposal selector and helps it address the scenarios with inaccurate temperature proposals.

The temperature proposal selector module is made of a neural network that consists of eight layers. The selector firstly takes the historical system state and the historical control actions $\langle$ $(S_{t-k}, A_{t-k})$, $(S_{t-k+1}, A_{t-k+1})$, $\cdots$, $(S_{t-1}, A_{t-1})$ $\rangle$ as the part of the network input. Then this historical information will be sent to the first network layer. Including the first layer, there are four sequentially connected one-dimensional convolutional layers with the ReLU activation function on the bottom of the network. The output feature of these layers is two-dimensional in each data sample, and we convert it to a one-dimension feature vector $F_1$. Then the rest of the network inputs are two selected temperature proposals, the raw sensor reading ${pl}_1$ and the temperature prediction value ${pl}_2$, and they will be concatenated with the feature vector $F_1$. As shown in Fig.~\ref{fig:overview}, four fully-connected layers receive features vector $F_1$ and with those two selected temperature proposals ${pl}_1, {pl}_2$  (note that the first three of them have RuLU activation function). The last fully-connected layer has two neurons, which will be sent to a softmax layer and output a binary classification result by selecting the index with the maximum output value. 

Furthermore, the construction of the training data used for the temperature proposal selector differs from the previous module. The historical system state $S_{t-i} (i \in [1,k])$ and the historical control actions $A_{t-i} (i \in [1,k])$ are selected from the simulation data which is the same as in Section~\ref{temperature predictor}. The data in the two temperature proposals contain both normal and faulty data. So the training data consists of three types:

\begin{myitemize}
    \item Training data: $\langle$ historical system states $S_{t-i}$, control actions $A_{t-i}$, $(i \in [1,k])$, normal temperature, faulty temperature $\rangle$. \\
    Label: ($1, 0$).
    
    \item Training data: $\langle$ historical system states $S_{t-i}$, control actions $A_{t-i}$, $(i \in [1,k])$, faulty temperature, normal temperature $\rangle$. \\
    Label: ($0, 1$).
    
    \item Training data: $\langle$ historical system states $S_{t-i}$, control actions $A_{t-i}$, $(i \in [1,k])$, faulty temperature, faulty temperature $\rangle$. \\
    Label: $1$ is assigned to the value that is closer to the normal temperature. The other is assigned with $0$.
\end{myitemize}

Similar to the data construction strategy in the temperature predictor module, the historical system states we utilize include the faulty sensor readings. Specifically, for enhancing the robustness of the temperature proposal selector, we use the historical system states under the independent and identically distributed (IID) faults with occurring probability $P_{sel}$. 
Besides, during constructing these data-label pairs, we sample the faulty temperature three times for each normal temperature value in the first and second kind of data-label pair. For the last kind of data-label pair, we sample the faulty temperature data four times for each historical sequence. All faulty temperature readings come from the IID faults. 
Finally, we learn the temperature proposal selector network through the cross-entropy loss function. The learning rate $lr_{sel}$ and training epochs $l_{sel}$ are set as in Table~\ref{hyperparam} later.

\begin{figure*}[tb]
\centering
\includegraphics[width=17cm]{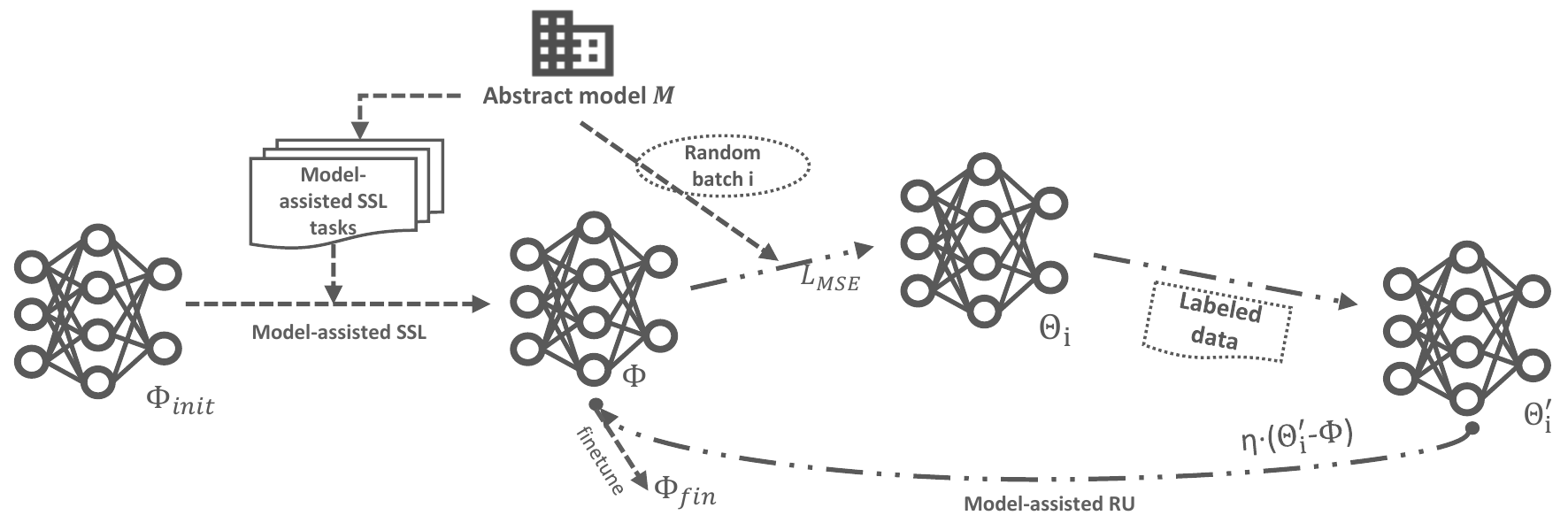}
\caption{Overview of our model-assisted learning for training with a limited amount of labeled data and an abstract physical model, where the algorithm consists of two stages -- model-assisted self-supervised learning (model-assisted SSL) and model-assisted redirected updating (model-assisted RU). The former stage creates auxiliary learning tasks from the abstract model, and the latter stage extracts knowledge leveraging the random batch from the physical model and explores a better updating direction. Then we get the final model through fine-tuning based on the pre-trained model from the previous two stages.
}
\vspace{-0.2cm}
\label{fig:model-assisted-learning}
\end{figure*}

\subsubsection{DRL-based controller for building HVAC system}
\label{DRL}
Because the thermal zone temperature in the next time step only relies on the observation of the current system state, the building HVAC control can be treated as a Markov decision process. 
We use a DQN-based DRL method that takes the current state $S^{DRL}_t$ as inputs, which contain
\begin{myitemize}
    \item Current physical time $t$,
    \item Current indoor air temperature $T^{in}_t$,
    \item Current ambient air temperature $T^{out}_t$,
    \item Current solar irradiance intensity ${Sun}_t$,
    \item Weather forecast in the next three time steps.
\end{myitemize}
The weather forecast includes ambient temperature and solar irradiance intensity $T^{out}_{t+1}, \cdots, T^{out}_{t+3}, {Sun}_{t+1}, \cdots, {Sun}_{t+3}$, which helps the network capture the trend of the environment. The deep Q-network $Q$ provides the Q-value estimation of current control actions. The algorithm takes the control action with the maximum Q-value and sends it to the HVAC system.

Furthermore, the goal of this DRL controller is to minimize total energy cost while maintaining indoor temperature within a comfort temperature bound $[T_l, T_u]$. The reward function $R_t$ collected from the control steps is designed accordingly as 

\begin{equation}
\label{eq3}
R_t = \alpha \cdot {R_c} + \beta \cdot {R_v}
\end{equation}
\begin{equation}
{R_c} = - cost(t-1, A_{t-1})
\end{equation}
\begin{equation}
{R_v} = -\sum_{i=1}^{n}{\max({T_l} - T^{in(i)}_t, 0) + \max(T^{in(i)}_t - {T_u}, 0)}
\end{equation}
where $\alpha$ and $\beta$ are the scaling factors. $R_c$ is the reward of energy cost, $R_v$ is the reward of temperature violation with respect to comfort temperature bound $[T_l, T_u]$. $cost(t-1, A_{t-1})$ is a price function that gives the money cost of the HVAC system from control time $t-1$ to $t$ under control action $A_{t-1}$. It is designed based on the local electricity price. Following the definition of the reward function, the update of deep Q-network is defined as

\begin{equation}
\begin{aligned}
    & Q_{t+1}(S^{DRL}_t, A_t) = Q_{t}(S^{DRL}_t, A_t) + \eta_0 (R_{t+1} \\
    & + \gamma \max_{A_{t+1}} Q_{t}(S^{DRL}_{t+1}, A_{t+1}) - Q_{t}(S^{DRL}_t, A_t)))
\end{aligned}
\end{equation}
where $\eta_0$ is the learning rate for the deep Q-network, and $\gamma$ is the decay factor of the accumulative reward.

\subsection{Model-assisted learning}
\label{model-assisted learning}

Our sensor fault-tolerant framework has three modules that require neural network training. The performance of a learning model is typically strongly correlated with the amount of available labeled data. 
However, collecting labeled data from real building operations takes significant amount of time, which often leads to the problem of training data insufficiency. With the techniques in~\cite{chen2020transfer, lissa2020transfer}, the training time and the required data of the DRL control module can be substantially reduced. With the special training data construction strategy introduced in Section~\ref{temperature proposal selector}, the selector also has sufficient data for training. Thus, we focus our effort on the data insufficiency issue for the temperature predictor. We develop a novel model-assisted learning method to combine a limited number of accurately-labeled data $D^{L}$ with the knowledge we can gain from an abstract physical model $M$ for the training, as shown in Fig.~\ref{fig:model-assisted-learning}. 


Our model-assisted learning consists of two stages:  model-assisted self-supervised learning (called model-assisted SSL) and model-assisted redirected updating (called model-assisted RU). To begin with, we realize that the biggest challenge in this learning scenario is that we do not have enough training data (even unlabeled data), which makes the typical semi-supervised or weakly supervised learning methods not applicable. However, one available resource that we can leverage is the human-designed abstract physical models for buildings. While they may not accurately describe the building dynamics, they do reflect some of the fundamental physical laws for the system. By `extracting' these physical laws, we can significantly improve the learning process and reduce the need for training data.
Specifically, for each element $u$ in the neural network input $s$, we can define its range based on its physical meaning. Then considering the range for all the elements in $s$, we can define a space $H$ that contains all $s$ in its range combinations and $s \in H$. Note that $H$ is a space that is much larger than the actual data distribution for network inputs, which means that many unrealistic cases that will never happen in the real world might still occur when sampling from $H$.  

In model-assisted learning process, a required step is to collect enough samples from data space $H$. However, we notice that the input size of the neural network (temperature predictor), $(2+2n)k$, is large. Taking $n=4, k=20$ for example, the sampling is on a 200-dimensional continuous data space, which is too expensive for simple uniform sampling. Thus, we only sample the first historical state uniformly among that sub-space of size $2+2n$, and then feed that historical state to the physical model $M$ to predict the next historical state. Then we generate the latter historical states by repetitively applying the previous historical states to the physical model. In this way, we can collect the sample sequences of length $k$ and form an input data set $D$. We then divide $D$ into mini-batches and call them random batches $\{\mathbf{x} | \mathbf{x} \subset D\}$, and we denote the batch size of $\mathbf{x}$ as $b$. With the random batch, we can design the steps in model-assisted SSL and model-assisted RU.

In the first stage of model-assisted SSL, we aim to construct auxiliary learning tasks from the abstract model $M$ to decide an pre-trained weights for the neural network. The sampled data $d \in \mathbf{x} \subset D$ is a simulated states sequence based on the abstract physical model $M$, and the time length of the sequence is $k$. And we create $k$ auxiliary learning tasks based on the input sequence $d$. Specifically, for auxiliary task $i$, it is a regression task. The corresponding training data is $\{(d_i, y_i) | d_i$ equals to $d$ except that the indoor temperature in $d$ at time step $i$ is set to $-1$, $y_i$ is the value of indoor temperature in $d$ at time step $i$, $d \in \mathbf{x} \subset D\}$. In other word, we try to predict the missing state generated by the abstract model. The training step last for $l_{{MS}_i}$ epochs with batch size as $b_{MS}$ and learning rate as $\eta_0$. In auxiliary task $i$, we also need to edit the original neural network with some changes. We keep the first three fully-connected layers but add two extra fully-connected layers (individually for each task $i$) following the third layer. The newly added layers will provide the output for task $i$. This means that we share the feature extraction layers among all the auxiliary learning tasks, and those tasks will help the neural network leverage the relation of variables in the states sequence for constructing pre-trained weights. The model-assisted SSL will be conducted for $l_{MS}$ epochs, and we start from the randomly initialized neural network weights $\Phi_{init}$. The auxiliary learning tasks are run in order from tasks $1$ to $k$ in each epoch, and then we get a pre-trained weight $\Phi$ for our next stage.

In the second stage of model-assisted RU, we target on redirecting the updating direction when extracting the knowledge from the abstract model. In each update step $i$, we start from the current network weights $\Phi_i$ (the initial weights in this stage is $\Phi_0$ = $\Phi$), and select a random batch $\mathbf{x}$ and apply the abstract model $M$ on them to get the corresponding labels $\mathbf{y}$. Next, we are able to get a new model $\Theta_i$ by updating the parameters on $\Phi_i$ using the random batch $\mathbf{x}$ and its corresponding labels $\mathbf{y}$, which follows the equation
\begin{equation}\label{eq_distill}
    \Theta_i = \Phi_i - \eta_2 \nabla_{\Phi_i} \mathcal{L}_{MSE}(\Phi_i),
\end{equation}
where $\mathcal{L}_{MSE}$ is the mean square error loss and $\eta_2$ is the learning rate. The training lasts for $n_{iter}$ iterations, and uses a new sampling data batch for each iteration.

Next, we employ accurately labeled data $D^L$ to further fine-tune the model $\Theta_i$ from the last step by $l_{ft}$ epochs, and update to the model weights $\Theta_i^{'}$, as described in the following equation
\begin{equation}
    \Theta_i^{'} = \Theta_i - \eta_3 \nabla_{\Theta_i} \mathcal{L}_{target}(\Theta_i),
\end{equation}
where $\mathcal{L}_{target}$ is the loss function for the target task and $\eta_3$ is the learning rate for this step.

Looking back to what we have done in this stage, we first use the random batch $\mathbf{x}$ to distill the physical model $M$ as a further pre-trained model for the current step, and then we fine-tune the model using the accurately labeled data. The final performance of model $\Theta_i^{'}$ should reflect the quality of the initial update from $\Phi_i$ to $\Theta_i$, which depends on the corresponding random batch $\mathbf{x}$ and the abstract model $M$'s output knowledge $\mathbf{y}$. $\mathcal{L}_{target}$ shows a reference value considering the improvement brought by the Equation~\eqref{eq_distill}, while $\Theta_i^{'} - \Phi_i$ provides a better updating direction for the current knowledge extraction step compared to the  Equation~\eqref{eq_distill}. Thus, we determine the true updating step for the initial model $\Phi_i$ as
\begin{equation}
    \Phi_i = \Phi_i - \eta_1(\Theta_i^{'} - \Phi_i)
\end{equation}
Following this updating steps for $n_{iter}$ iterations, we then use all the accurately labeled data $D^{L}$ to fine-tune the extracted model $\Phi_i$ to achieve our target model $\Phi_{fin}$. The fine-tuning step has the learning rate $\eta_3$ by $l_{ft}$ epochs.

%% file: section/experiment.tex
\section{Experimental Results}
\label{experiment}


\subsection{Fault patterns, metrics and physical model}
\label{Fault patterns and Metrics}

\paragraph{Fault patterns:} We consider two types of fault patterns for the sensors in every thermal zone in our experiments. Both patterns could be caused by passive faults or cyber attacks.
\begin{myitemize}
    \item In the first type of faulty sensor readings, we postulate that the fault happens at each time step with a probability $p_1$. Note that the fault can happen in each simulation step, not only on the control steps. If the fault occurs, it uniformly selects a random number from $[T^{out}_l, T^{out}_u]$ (which is the upper and lower boundary of the ambient temperature in our experiments) to replace the original sensor temperature reading.  We call this type of faults the IID faults because they have the same probability, same distribution, and independent at each time step.
    \item For the second type of faulty sensor reading, the fault happens at each time step with a probability $p_2$. The difference between it and the former one is that the second fault will last for $\varpi$ simulation steps and not always happens individually among the time period. Thus this type of fault can cause larger damage to the system than the first one. And we call it continuous faults.
\end{myitemize} 

\paragraph{Metrics for evaluation:} We evaluate the sensor fault-tolerant temperature control results based on the average indoor temperature violation rate $\theta_i$ for each thermal zone $i$ and the total energy cost for running the HVAC system. We evaluate the performance of model-assisted learning on the temperature prediction task with a four-zone building. The measurement for the predictor is based on the root mean square error (RMSE) between the prediction and the actual temperature value.

\paragraph{Abstract physical model:}
\label{physical model} Here we introduce the abstract physical model we used in experiments for model-assisted learning. The mass and energy conservation law for a building zone is presented in Equation~\eqref{eqn-zone-differential}, where the left of the equation represents energy changes in the zone, the first term at the right represents the introduced HVAC energy to the zone, and the second term at the right is the thermal load in the zone.
The thermal load $\dot q_l$ is related to many building system and control parameters such as envelope constructions, internal heat gains, zone air temperature setpoints, etc., which eventually leads to a nonlinear differential equation to solve.
For simplification, an abstract model for the zone air temperature dynamics is derived as in Equation~\eqref{eqn-zone-narx}. 
This model explicitly relates zone air temperature to system thermal inertia (e.g., historical zone air temperatures), zonal supply air mass flowrate $\dot m$, outdoor air temperature $T^{out}$ and estimated modeling error term $e$. 
$\hat T$ and $T$ are the predicted and measured temperature, respectively.
$m$ is the zone air thermal mass.
$\dot m$ is the zonal supply air mass flowrate.
$C_p$ is the zone air specific heat.
$e$ represents an error term.
Superscripts $sa$, and $out$ are the supply air, and outdoor air, respectively.
$\alpha$, $\beta$, and $\gamma$ are identified coefficients observed from the given short-term historical data.

\begin{equation}\label{eqn-zone-differential}
    mC_p\frac{dT}{dt} = \dot mC_p(T^{sa}-T) + \dot q^l 
\end{equation}
\begin{equation}\label{eqn-zone-narx}
    \hat T_{t+1} = \alpha T_{t} + \beta\dot m_{t+1} + \gamma \hat T^{out}_{t+1} + e_{t+1} 
\end{equation}
\begin{equation}\label{eqn-zone-narx-err}
    e_{t+1} = \sum_{j=0}^{L-1} \frac{\hat T_{t-j}-T_{t-j}}{L}
\end{equation}

\subsection{Experiment settings}

\begin{figure}[htb]
\centering
\includegraphics[width=8cm]{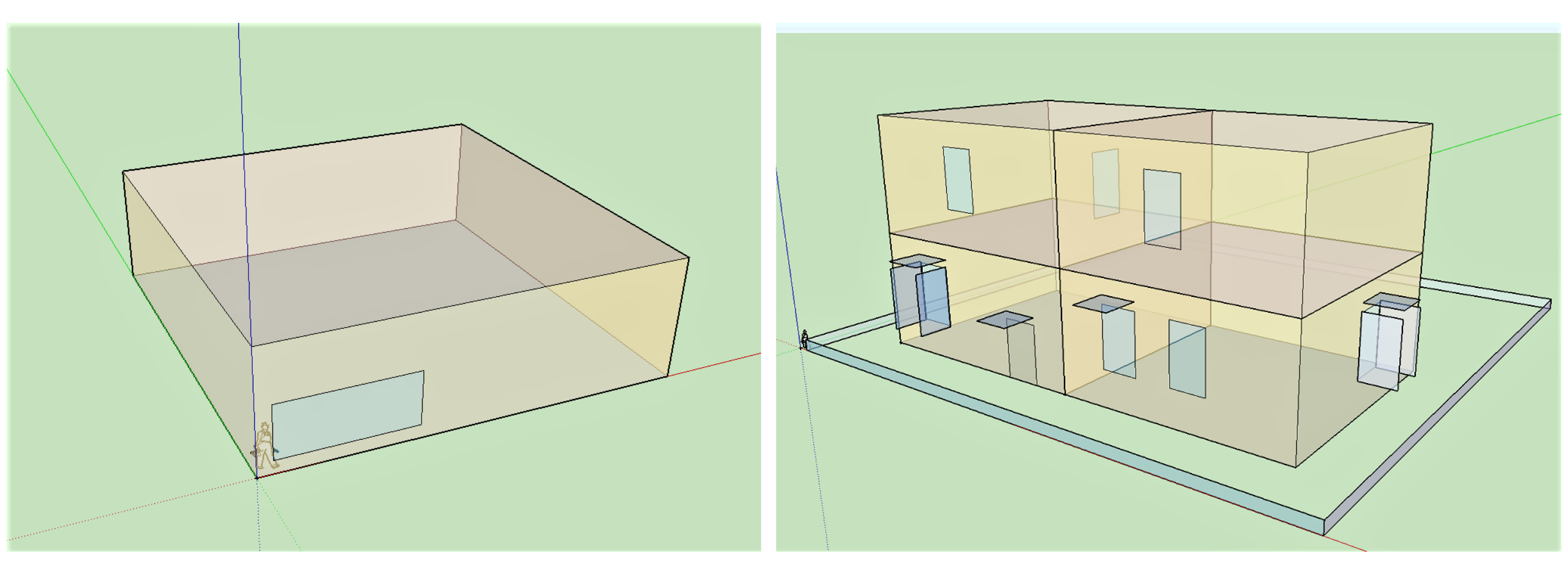}
\caption{Rendering of the experimental buildings.}
\label{building_vis}
\end{figure}

The experiments are run on an Ubuntu OS server equipped with NVIDIA TITAN RTX GPU cards. The learning algorithm implementations are based on the Pytorch framework. The Adam optimizer~\cite{kingma2014adam} is utilized for all neural networks' training. We use the EnergyPlus~\cite{crawley2000energy} simulation tool to simulate the behavior of real buildings. Note that this is only for experimentation purpose. In practice, our tool will be deployed directly on real buildings with the modules trained on the data collected from those buildings.
Moreover, the interaction between the building simulations in EnergyPlus and the Pytorch learning algorithms is implemented through the Building Controls Virtual Test Bed (BCVTB)~\cite{wetter2011co}. We use a single-zone building and a 4-zone building as the target buildings for conducting our experiments, which are visualized in Fig.~\ref{building_vis}. The building simulation utilizes the summer weather data in August at Riverside, California, USA, which is obtained from the Typical Meteorological Year 3 database~\cite{wilcox2008users}. The hyper-parameter settings mentioned in the previous sections are shown in Table~\ref{hyperparam}.

\begin{table}
\newcommand{\tabincell}[2]{\begin{tabular}{@{}#1@{}}#2\end{tabular}}
\begin{tabular}{c|c|c|c}  
\hline  
Parameter & Value & Parameter & Value  \\  
\hline  
\tabincell{c}{Temperature- \\proposal- \\selector layers} & \tabincell{c}{[2+2$n$,512,256,\\256, 128,256,\\256,256,2$n$]} &\tabincell{c}{DQN layers \\ \  \\$T_{l}$} & \tabincell{c}{[9+$n$,50,100,\\200,400,16] \\ 19 \textcelsius}\\
\tabincell{c}{Predictor-\\ layers} & \tabincell{c}{[(2+2$n$)$k$,512,256,\\256,256,256,$n$]}  & \tabincell{c}{ $T_{u}$ \\ $P_{pre}$} & \tabincell{c}{24 \textcelsius \\ 0.1} \\
$T^{pre}_{ofs}$ & 22 & $P_{sel}$ & 0.3 \\
$l_{ft}$ & 3 & $\alpha$ &  1e-3 \\
$\beta$ & 6.25e-4 & $b_{MS}$ & 40 \\
$\eta_0$ & 1e-3 & $\eta_1$ &  1e-4 \\
$\eta_2$ & 1e-6 & $\eta_3$ & 1e-3  \\
$L$ & 5760 & $k$ & 20\\
$\Delta t_s$ & 1 min & $\Delta t_c$ & 15 min \\
$T^{out}_l$ & 10 \textcelsius & $T^{out}_u$ & 40 \textcelsius \\
$v$ & 2 & $lr_{sel}$ & 1e-4\\
$l_{sel}$ & 50 & $T^{air}$ & 10 \textcelsius\\
$m$ & 2 & $\eta_0$ & 0.003\\
$\gamma$ & 0.99 & $b$ & 32 \\
$l_{{MS}_i}$ & 3 & $l_{MS}$ & 2 \\
\hline

\hline
\end{tabular}  

\caption{Hyper-parameters used in our experiments.}
\label{hyperparam}
\end{table}

\subsection{Evaluation of sensor fault-tolerant framework on IID and continuous faults}
This section shows the performance of our sensor fault-tolerant framework and its comparison with a standard DQN controller. The experiments are conducted on a single-zone building and a four-zone building under different sensor fault patterns.

\subsubsection{Against IID faults}
We first study how much the sensor fault-tolerant framework can protect the control performance from the IID faults. The IID faults happen individually at each simulation step with the probability $p_1$, and we test the case where $p_1$ is chosen from $[0, 0.1, 0.2, 0.4, 0.6, 0.8]$. The model is first tested on a single zone building. Table~\ref{ft-iid-1zone} shows the results comparison between the standard DQN controller (DQN) and our sensor fault-tolerant framework (FTF). We can see that the typical DQN controller's performance significantly deteriorates when facing the IID faults, as the heavily faulty sensor data nearly paralyzed the normal function of the neural network. The problem gets worse with the fault occurring probability $p_1$ becomes larger. 
For our sensor fault-tolerant framework, the average temperature violation rate remains very low under varying degree of IID faults ($86.4\%$ to $98.2\%$ reduction in violation rate when compared with standard DQN under fault probability from 0.1 to 0.8). Moreover, even with our approach's much more robust control, the energy cost does not increase much compared to the non-faulty case, which shows the cost-effectiveness of our sensor fault-tolerant approach.

\begin{figure}[htb]
\centering
\includegraphics[width=9.3cm]{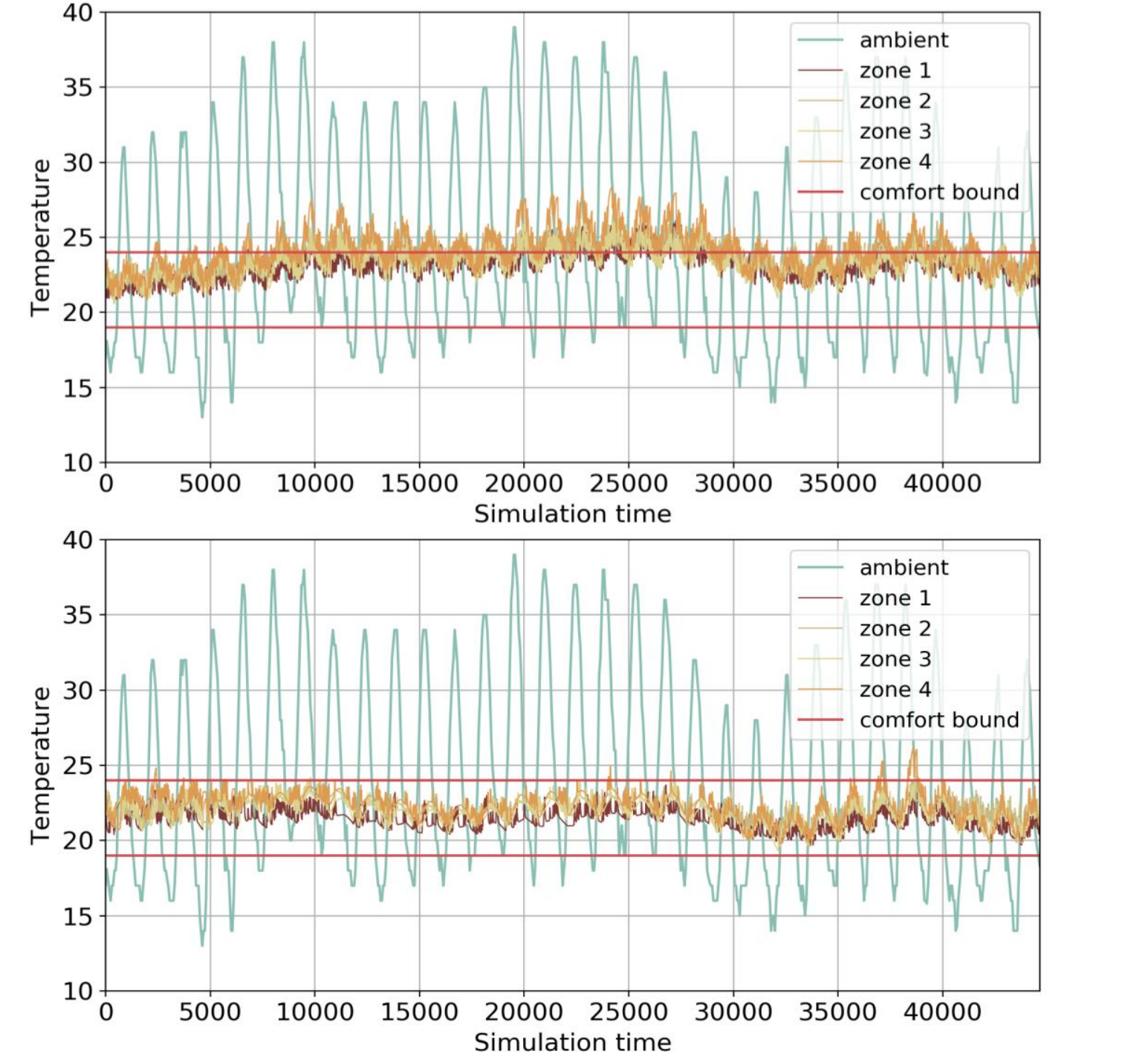}
\caption{4-zone building temperature under IID faults with $p_1 = 0.4$ without FTF (above) and with FTF control (below).}
\label{ftf_vis}
\end{figure}

We also tested our framework on a 4-zone building against the IID faults, and Table~\ref{ft-iid-4zone} shows its comparison with the standard DQN. $\theta_1$ to $\theta_4$ are the temperature violation rate for each of the 4 thermal zones. Again, we can clearly see that our approach can maintain the violation rate at a low level under varying level of sensor faults, and can significantly outperform the standard DQN ($84.89\%$ to $97.45\%$ reduction in violation rate under fault probability from 0.1 to 0.8). It is also worth mentioning that when there is no fault, our framework will not introduce additional overhead. Finally, Fig.~\ref{ftf_vis} also provides a visualization of the temperature change on the 4-zone building under IID faults with $p_1 = 0.4$ with/without the sensor fault-tolerant framework, and we can clearly see the effectiveness of our framework in keeping the temperature within comfort bound under faults.


\subsubsection{Against continuous faults}
We then evaluate our approach against continuous faults. Similar to what we have shown in the previous section, the model is tested on a single-zone building and a four-zone building, with the probability $p_2$ set to $0.1$ and $\varpi$ selected from $0$ to $5$. The comparison between our approach and the standard DQN is presented in Table~\ref{ft-con-1zone} and Table~\ref{ft-con-4zone}. The temperature violation rates in the tables are all higher than the previous section under the same fault probability, which indicates that the continuous faults can cause more damage than the IID faults. As shown in the table, the standard DQN controller drastically increases the violation rate for 13$\times$ to 51$\times$ for single zone and 6$\times$ to 39$\times$ for four-zone under continuous faults. In comparison, our approach can effectively maintain the violation rate at a low level ($55.1\%$ to $86.4\%$ reduction for single zone and $70.0\%$ to $87.5\%$ reduction for four-zone in violation rate when compared with the standard DQN under fault probability from 0.1 to 0.8). 


\begin{table}

\centering
\vspace{-0.1cm}
\begin{tabular*}{8.8cm}{l|lllllll}
\hline
\multicolumn{1}{l|}{}  & $p_1$ &  0 & $0.1$ & $0.2$ & $0.4$ & $0.6$ & $0.8$ \\ \hline
\multirow{2}{*}{DQN}   & $\theta$ &  0.08   &   1.18    &   2.18    &   3.59    &  5.90  &  11.19  \\
                       & Cost&  250.03   &   245.79    &   239.74    &   235.73    &  228.42 &  223.22  \\ \hline
\multirow{2}{*}{FTF}   & $\theta$ &  0.15   &   0.16    &   0.45    &   0.20    &  0.62  &  0.20 \\
                       & Cost&     250.97    &   251.14    &   250.18    &   254.87  & 259.40 & 270.08 \\ \hline

\hline
\end{tabular*}
\caption{Comparison between standard DQN controller and our sensor fault-tolerant framework (FTF) on a single-zone building under IID faults. $p_1$ is the fault occurring probability. $\theta$ is the average indoor temperature violation rate (\%).}
\vspace{-0.3cm}
\label{ft-iid-1zone}
\end{table}

\begin{table}  

\centering
\begin{tabular*}{8.8cm}{c|lllllll}
\hline
\multicolumn{1}{l|}{}  & $p_1$ &  0 & $0.1$ & $0.2$ & $0.4$ & $0.6$ & $0.8$ \\ \hline
\multirow{5}{*}{DQN}   & $\theta_1$ &  0.0   &   1.68    &   3.76    &   14.64    &  32.18  & 48.47  \\
                       & $\theta_2$ &  0.37   &   6.27    &   16.3    &   30.6    &  50.70  & 59.56  \\
                       & $\theta_3$ &   1.16  &   3.17    &   7.59    &   15.5    &  27.22  & 36.4  \\
                       & $\theta_4$ &   1.42  &   9.33    &   18.46    &   28.22    &  43.50  &  48.13 \\
                       & Cost &  258.33   &   246.07    &   235.17    &    220.53   &  197.86  & 187.76  \\ \hline
\multirow{5}{*}{FTF}   & $\theta_1$ &  0.07   &   0.02    &   0.09    &   0.07    &  0.01  & 0.00  \\
                       & $\theta_2$ &  0.34   &   0.32    &   0.38    &   0.13    &  0.02  & 0.11  \\
                       & $\theta_3$ &  1.16   &   0.70    &   0.60    &   0.44    &  0.64  &  0.74 \\
                       & $\theta_4$ &  1.59   &   1.51    &   1.91    &   2.63    &  2.86  &  4.06 \\ 
                       & Cost &   258.8   &    259.11   &   265.79    &    301.50   &  322.64  & 321.17  \\ 
                       \hline
                       
                       \hline
\end{tabular*}
\caption{Comparison between standard DQN controller and our sensor fault-tolerant framework (FTF) on a four-zone building under IID faults. $p_1$ is the fault probability. $\theta_i$ is the avg. indoor temperature violation rate (\%) in thermal zone $i$.}
\label{ft-iid-4zone}
\end{table}

\begin{table}  
\centering
\begin{tabular*}{8.8cm}{l|lllllll}
\hline
\multicolumn{1}{l|}{}  & $\varpi$& $0$ & $1$ & $2$ & $3$ & $4$ & $5$  \\ \hline
\multirow{2}{*}{DQN}   & $\theta$ &  0.08   &   1.18    &   2.35    &   3.01    &  3.16  &  4.17  \\
                       & Cost&   250.03  &   245.79    &   237.95    &   236.48    &  237.67 &  231.25  \\ \hline
\multirow{2}{*}{FTF}   & $\theta$ &  0.15   &   0.16    &    1.43    &   1.65    & 1.10  & 1.87  \\
                       & Cost&  251.38   &    260.19   &   244.12    &    249.95   &  253.37   & 252.30 \\ \hline

\hline
\end{tabular*}
\caption{Comparison between standard DQN controller and our sensor fault-tolerant framework (FTF) on a single-zone building under continuous faults. The fault lasts for $\varpi$ steps. $\theta$ is the avg. indoor temperature violation rate (\%).}
\vspace{-0.5cm}
\label{ft-con-1zone}
\end{table}

\begin{table}  
\centering
\begin{tabular*}{8.8cm}{c|lllllll}
\hline
\multicolumn{1}{l|}{}  & $\varpi$& $0$ & $1$ & $2$ & $3$ & $4$ & $5$  \\ \hline
\multirow{5}{*}{DQN}   & $\theta_1$ &  0   &    1.68   &   5.27    &   9.68    &  14.41  &  18.08 \\
                       & $\theta_2$ &  0.37   &  6.27     &  13.26     &   21.67    & 29.57   & 33.53  \\
                       & $\theta_3$ &  1.16   &   3.17    &   11.12    &   14.77    &  20.05  &  22.58 \\
                       & $\theta_4$ &  1.42   &   9.33    &   24.54    &   34.98    &  39.54  & 43.19  \\
                       & Cost &  258.33   &   246.07    &   231.42    &   219.10    &  213.99  &  211.20 \\ \hline
\multirow{5}{*}{FTF}   & $\theta_1$ &  0.07   &   0.02    &   0.11    &   0.35    &  0.90  &  1.12 \\
                       & $\theta_2$ &  0.34   &   0.32    &    2.62   &   3.91    &  5.88  & 10.61  \\
                       & $\theta_3$ &  1.16   &   0.70    &   1.78    &   2.79    &  3.94  & 5.35  \\
                       & $\theta_4$ &  1.59   &   1.51    &   4.33    &   8.04    &  12.6  & 18.09  \\ 
                       & Cost &  258.86   &   259.11    &   269.97    &   268.97    &  265.18  & 261.26  \\ 
                       \hline
                       
                       \hline
\end{tabular*}
\caption{Comparison between standard DQN controller and our sensor fault-tolerant framework (FTF) on a four-zone building under continuous faults. The fault lasts for $\varpi$ steps. $\theta_i$ is the avg. indoor temperature violation rate (\%) in thermal zone $i$.}
\vspace{-0.2cm}
\label{ft-con-4zone}
\end{table}

\subsection{Evaluation of model-assisted learning}
\label{mal-exper}
In this section, we conduct experiments on the model-assisted learning algorithm and demonstrate its improvement in the performance of the temperature predictor module. Note that the data only contains non-faulty data in this section for avoiding other factors that may affect the evaluation, which means that there is no sensor fault in both training and testing.

\begin{table}  
\centering
\vspace{-0.0cm}
\begin{tabular*}{8.5cm}{lccccc}  
\hline  
Amount of data & 360 & 720 & 1440 & 2880 & 5760\\  
\hline
\textit{Labeled data only} & 0.650 & 0.447 & 0.265 & 0.198 & 0.137\\
\textit{Distillation + fine-tuning} & 0.649 & 0.412 & 0.258 & 0.222 & 0.149\\
\textit{Model-assisted SSL} & 0.354 & 0.234 & 0.178 & 0.114 & 0.094\\
\textit{Model-assisted RU} & 0.351 & 0.270 & 0.200 & 0.146 & 0.081\\
\textit{Model-assisted learning} & 0.261 & 0.226 & 0.130 & 0.052 & 0.045\\
\hline  

\hline  
\end{tabular*}
\caption{Comparison of different learning strategies on temperature predictor performance. The first line shows training with labeled data only. The second line shows the distillation approach as in~\cite{hinton2015distilling}. The third line shows using only the first stage (model-assisted SSL) of our model-assisted learning approach, and the fourth line shows only using the second stage (model-assisted RU). The last line shows using both stages, i.e., our model-assisted learning approach. 
}
\vspace{-0.4cm}
\label{mal-regression}
\end{table}

\begin{figure}[htb]
\centering
\includegraphics[width=7.7cm]{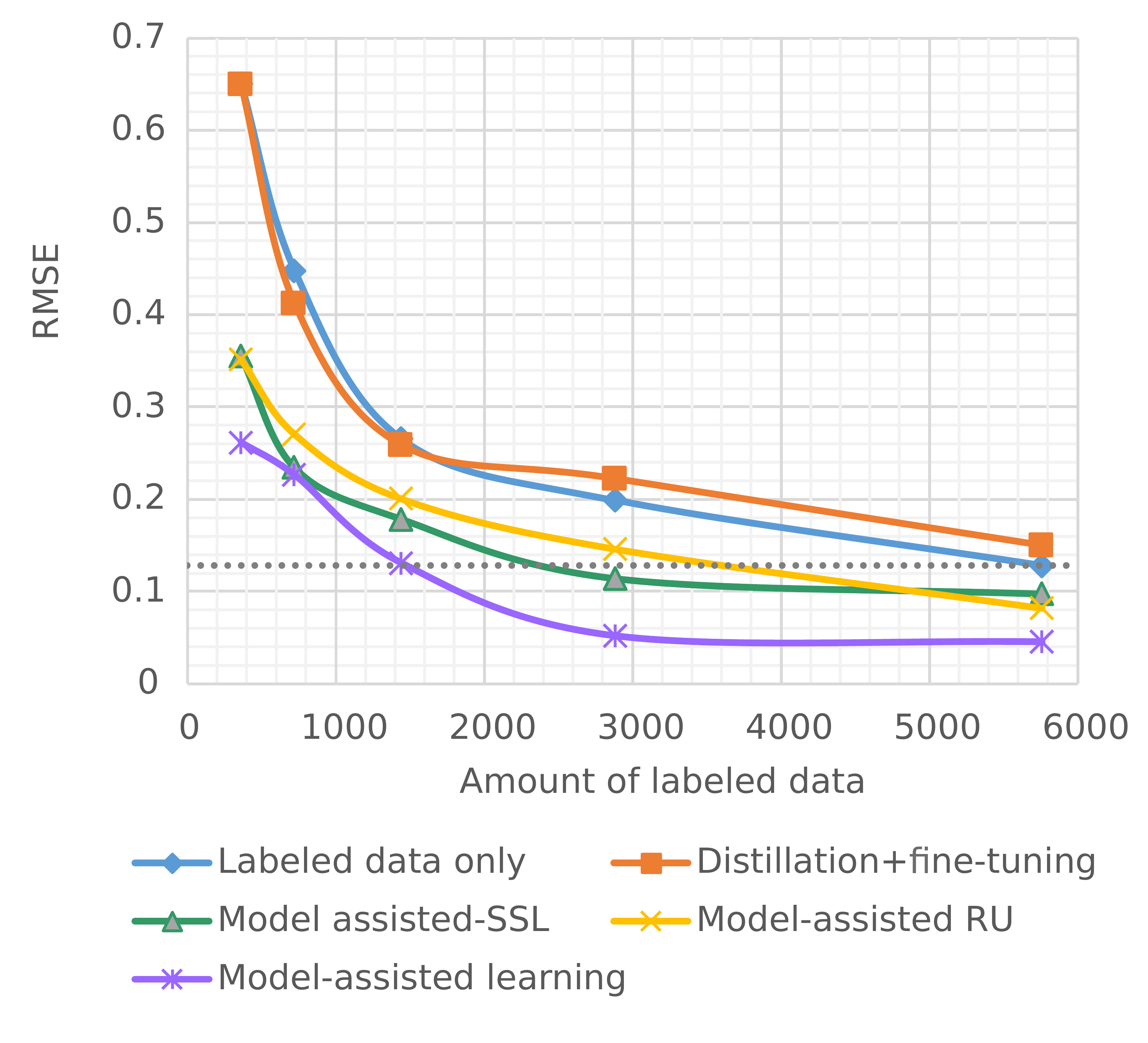}
\vspace{-0.2cm}
\caption{Comparison of different learning strategies on temperature prediction performance, including \textit{Labeled data only} (blue line), \textit{Distillation+fine-tuning} (orange line), \textit{Model-assisted SSL} and \textit{RU} (green $\&$ yellow line), \textit{Model-assisted learning} (purple line). We can observe from the figure that \textit{Model-assisted learning} only requires around 1400 data samples to reach the RMSE of using \textit{Labeled data only} with 5760 samples, i.e., only needs 1/4 of the labeled data by leveraging the abstract physical model via our approach.}
\vspace{-0.3cm}
\label{model-assist}
\end{figure}

We employ an abstract physical model introduced in Section~\ref{physical model} for a four-zone building. The abstract model itself has the temperature prediction value with RMSE at 0.832. Then if we only use the accurately labeled data collected from the building to train the neural networks in the temperature predictor module, which is shown in the first line in Table~\ref{mal-regression} (the model named \textit{Labeled data only}), we can see that the RMSE remains at the relatively high level (note that the average indoor temperature change between two time steps is around 0 to 0.25), e.g., 0.265 for 1440 data samples, and 0.198 for 2880 data samples. More labeled data leads to more accurate model prediction. The maximum amount of available data is 5760 samples for the simulation of eight days. 

In addition to model-assisted learning, we also test another idea for leveraging the abstract physical model $M$ to gain better performance, i.e., using the abstract physical model to set initial weights for a neural network, so the network may cost less training data for reaching higher accuracy as it searches from a better initial point. The related technique for obtaining this initial value is model distillation~\cite{hinton2015distilling}. However, as mentioned earlier, choosing the data to feed the neural network is challenging for distillation. Here we use the same sampling approach as proposed in Section~\ref{model-assisted learning}, i.e., sampling from data space $H$ and feeding the samples $\mathbf{x}$ to the abstract physical model $M$. Then we get the corresponding data pair ($\mathbf{x}, \mathbf{y}$), and train the network using ($\mathbf{x}, \mathbf{y}$) with learning rate $\eta_2$ for $n_{iter}$ iterations (a new sampling data batch for each iteration). Next, we fine-tune this newly trained model with learning rate $\eta_3$ in $l_{ft}$ epochs on the accurately labeled data. The model obtained in this way is named as \textit{Distillation + fine-tuning} (which is shown in the second line of the Table~\ref{mal-regression}).

Finally, we apply our proposed model-assisted learning to leverage the abstract physical model. To understand how much each stage contributes to the final performance, we add the results of only applying one of the two stages, which are the third line (\textit{Model-assisted SSL}) and fourth line (\textit{Model-assisted RU}) of Table~\ref{mal-regression}, respectively. And when combining both, the result is our \textit{Model-assisted learning}, as in the last line. 

We can observe from the table that, when the available sample is limited (360, 720, 1440), the building dynamics directly extracted by \textit{Distillation + fine-tuning} method can help reduce the RMSE. However, those extracted knowledge is only an inaccurate estimation, and the bias it brings prevents the model from achieving better result when there is more available labeled data (2880, 5760). On the other side, both stages in our \emph{Model-assisted learning} approach can make good use of the abstract model and reduce the RMSE among all cases. When combing the two together, with the same amount of labeled data, our \textit{Model-assisted learning} can achieve significantly better results than using only labeled data or distillation method. Such effectiveness is also visualized in Figure~\ref{model-assist} -- it plots the same results as Table~\ref{mal-regression}, but we can clearly see that for the same level of performance, our \textit{Model-assisted learning} approach only requires about 1/4 of the labeled data.


%% file: section/conclusion.tex
\section{Conclusion}
\label{conclusion}
In this paper, we present a novel learning-based sensor fault-tolerant control framework for building HVAC systems against faulty sensor readings, which includes neural network-based components for temperature prediction, temperature proposal selection, and DRL-based HVAC control. We also introduce a model-assisted learning approach that leverages abstract physical model to overcome the difficulty in training data insufficiency. Experimental results demonstrate the effectiveness of our framework and the model-assisted learning method.
